\documentclass[
    ,final            
  ]
  {aipproc}

\layoutstyle{8x11double}
\begin{document}

\title{Nucleosynthesis in Early Neutrino Driven Winds}

\author{R.D. Hoffman}{
  address={Lawrence Livermore National Laboratory, PO Box 808, L-414, Livermore, CA 94550 USA}
}

\author{J.L. Fisker}{
  address={Lawrence Livermore National Laboratory, PO Box 808, L-414, Livermore, CA 94550 USA}
}

\author{J. Pruet}{
  address={Lawrence Livermore National Laboratory, PO Box 808, L-059, Livermore, CA 94550 USA}
}

\author{S.E. Woosley}{
  address={Department of Astronomy \& Astrophysics, UC Santa Cruz, Santa Cruz, CA 95064 USA}
}

\author{H.-T. Janka}{
  address={Max Plank Institute for Astrophysics, Karl-Schwarzschild-Str. 1, 85741 Garching, Germany}
}

\author{R. Buras}{
  address={Max Plank Institute for Astrophysics, Karl-Schwarzschild-Str. 1, 85741 Garching, Germany}
}

\begin{abstract}
Two recent issues realted to nucleosynthesis in early proton-rich neutrino winds are 
investigated. In the first part we investigate the effect of nuclear physics uncertainties
on the synthesis of $^{92}$Mo and $^{94}$Mo. Based on recent experimental results, 
we find that the proton rich winds of the model investigated here can not be the only
source of the solar abundance of $^{92}$Mo and $^{94}$Mo. In the second part we investigate the 
nucleosynthesis from neutron rich bubbles and show that they do not contribute to 
the nucleosynthesis integrated over both neutron and proton-rich bubbles and proton-rich winds.
\end{abstract}

\maketitle

\section{Introduction}
Over the past decade improvements in neutrino-transport and multi-dimensional computer 
simulations have lead to a new understanding of the conditions that lead to nucleosynthesis 
of the elements above iron in core-collapse supernovae. 
Immediately following the bounce on the proto-neutron star, the shock fully 
photodisintegrates the infalling material turning it into electron--position pairs, 
neutrons, and protons. As the nascent neutron star continues to collapse it 
liberates $10^{53} \textrm{ergs}$ over the span of $\sim 10$ seconds primarily in 
the form of neutrinos. This enormous neutrino flux is deposited in the low density 
region of photodisintegrated matter inside the gain radius between the neutron star 
and the accretion shock of the still infalling material and heats it to temperatures 
in excess of 10 billion K while driving mass away in the form 
of a neutrino wind theoretically leading to the explosion of the supernova \cite{Qian96}. 
The strong flux of neutrinos and anti-neutrinos results in a detailed balance 
between protons and neutrons that favors the lighter mass protons depending 
on the respective neutrino spectra leading to an electron fraction that is proton-rich ($Y_e>0.5$)
\citep{Liebendoerfer03, Froehlich06b}. These protons and neutrons recombine into 
alpha particles that proceed via the 
$\alpha(\alpha n,\gamma)$ ${}^{9}\textrm{Be}(\alpha,n)$ ${}^{12}\textrm{C}$-reactions 
followed by a series of $(\alpha,\gamma)$-reactions or 
combined $(\alpha,p)(p,\gamma)$-reactions along $N=Z$ into the iron group, 
primarily ${}^{56}\textrm{Ni}$ and ${}^{60}\textrm{Zn}$ which form the seeds 
of the subsequent nucleosynthesis. 

From this point the resulting nucleosynthesis in the neutrino-driven wind essentially 
depends on the number of seed nuclei to the number of excess neutrons or protons that 
were frozen out and did not turn into seed nucleii ($Y_e$), the entropy per baryon, 
the expansion timescale of the ejecta and the amount of the ejecta. As the explosion 
evolves, an ejected mass element inherits some combination of these parameters and 
below $\sim 0.5\textrm{MeV}$ they remain fairly constant as the matter proceeds to freeze out. 

In this paper, we consider the early times when the wind still contains a proton 
excess because the rates for neutrino and positron captures on neutrons are faster 
than those for the inverse captures on protons. We consider two interesting problems 
which are discussed in the following two sections.  

\section{The puzzle of ${}^{92}\textrm{Mo}$}
The origin of $^{92}$Mo is a long standing puzzle of nucleosynthesis 
\citep[for reviews, see][]{Lambert92,Meyer94}. It is thought to originate in 
the proton-rich wind prior to the $r$-process in core collapse supernovae, 
but historically it has been underproduced in such models or subject to severe 
model constraints \citep{Fuller95,Hoffman96}. 

Recent supernova models show that the $Y_e\equiv\sum X_iZ_i/A_i$ of the innermost 
ejecta is greater that the $Y_e$ of the most abundant $p$-nuclei \citep{Qian96,Froehlich06b}. 
This implies the existence of surplus protons which allow the production of proton-rich 
$p$-nuclei nuclei by the $\nu rp$-process \citep{Pruet06}. However, similar to the 
$rp$-process in the X-ray burst scenario, there is an important waiting point at 
${}^{64}\textrm{Ge}$ which backs up material beyond the $t < 1\textrm{s}$ 
dynamic timescale of the innermost ejecta in core collapse \citep{Wallace81}. 

To solve this problem, it was suggested a new $\nu p$-process in which neutrinos convert 
some of the surplus protons into neutrons allowing the waiting points to be bridged via an 
$(n,p)$-reaction \cite{Froehlich06a}. This accelerates the flow into heavier elements and 
creates the light $p$-nuclei which are otherwise missing from the standard $r$-process. 
These calculations were independently confirmed by calculations based on simulations 
\cite{Pruet06,Janka03}. 

Still, relative to the solar abundances, both calculations show underproduction of 
${}^{92}\textrm{Mo}$ (the most abundant of the $p$-nuclei) relative to the $p$-nuclei of Ru and Pd. 
There are three possible reasons why ${}^{92}\textrm{Mo}$ is not co-produced with the 
other $p$-nuclei: 1) The $\nu p$-process is active, but ${}^{92}\textrm{Mo}$ is 
primarily synthesized at other sites. 2) The $\nu p$-process is not active, 
so another explanation is needed. 3) The $\nu p$-process is active, but the 
nuclear parameters that enter the nucleosynthesis calculation are incorrect.
In this paper, we investigate the third possibility. 

\subsection{The production of the light $p$-nuceli}

Nucleosynthesis results obtain from the sum total of the reaction flow in 
all the matter trajectories of the supernova ejecta. Here we only consider the 
reaction flow in ``trajectory 6'' (see Table 2 of \cite{Pruet06}) based on the 
model of \cite{Janka03} (see \cite{Rampp02} for specific code details and 
\cite{Pruet05} for more details). ``Trajectory 6'' is the trajectory where neutrino 
interactions are the most important in making the $p$-nuclei between Sr and Pd.

\begin{figure*}
\includegraphics[width=0.48\textwidth]{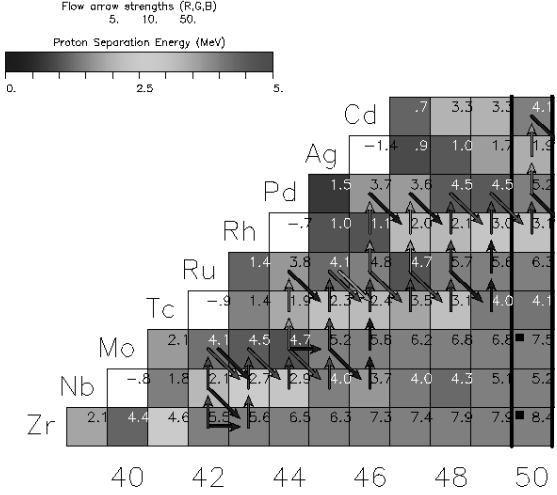}
\caption{A closeup of Figure 8 in \cite{Pruet06} for the region between 
Zr and Cd when ${\rm T}_9 = 2.06, \rho_5 = 2.74, {\rm and \ } 
Y_e = 0.561$ showng nuclear flows in the $A\sim 90$ region.
Each isotope is labled according to its proton separation energy
The arrows indicate the dominant net nuclear
flows. All net flows within a factor of 50 of the largest flow in this figure
($^{84}{\rm Nb}(p,\gamma )^{85}{\rm Tc} = 4.5\times 10^{-5} {\rm s}^{-1}$)
are shown.
The most important flows affecting $^{92,94}$Mo are
the proton capture flows on $^{92}{\rm Ru}$ and $^{93}{\rm Rh}$.}\label{fig:flow}
\end{figure*}

The $\nu p$-process starts on the iron group but it is halted at the 
long-lived ${}^{64}\textrm{Ge}$ waiting point which is known to be 
bridged by an (n,p)-reaction allowing the $\nu p$-process to continue 
\cite{Froehlich06a}.
The flow from ${}^{64}\textrm{Ge}$ passes through all even-even $T_z=(N-Z)/2=0$ 
isotopes until ${}^{88}\textrm{Ru}$ is reached \cite{Pruet06}. 
As fig.~\ref{fig:flow} shows, the pattern is broken because of the low 
proton separation energy of ${}^{90}\textrm{Ru}$ that prevents immediate proton captures 
up to ${}^{92}\textrm{Pd}$.
Instead the flow proceeds via 
${}^{90}\textrm{Ru}(n,p)$ ${}^{90}\textrm{Tc}(p,\gamma)$ ${}^{91}\textrm{Ru}$. 
A $(p,\gamma)$-reaction would result in the ${}^{92}\textrm{Rh}$ progenitor provided it 
does not get destroyed by another $(p,\gamma)$-reaction. Alternatively, 
an $(n,p)$-reaction to ${}^{91}\textrm{Tc}$ followed by a $(p,\gamma)$-reaction 
would result in the ${}^{92}\textrm{Ru}$ progenitor once again provided it does 
not get destroyed by another $(p,\gamma)$-reaction. In both cases the reverse 
reactions from  ${}^{93}\textrm{Pd}$ and ${}^{93}\textrm{Rh}$ would increase the survival 
of the $A=92$ progenitors.

Many of the relevant reaction rates, spins, partition functions, and proton 
separation are not known experimentally and the theoretical values are subject 
to considerable uncertainties which may change the flow. For instance, 
a 50\% yield increase in ${}^{92}\textrm{Mo}$ was found after a plausible 
$1\,\textrm{MeV}$ increase in the proton separation energy of 
${}^{91}\textrm{Ru}$ \cite{Pruet05}.

\begin{figure}
\includegraphics[width=0.48\textwidth]{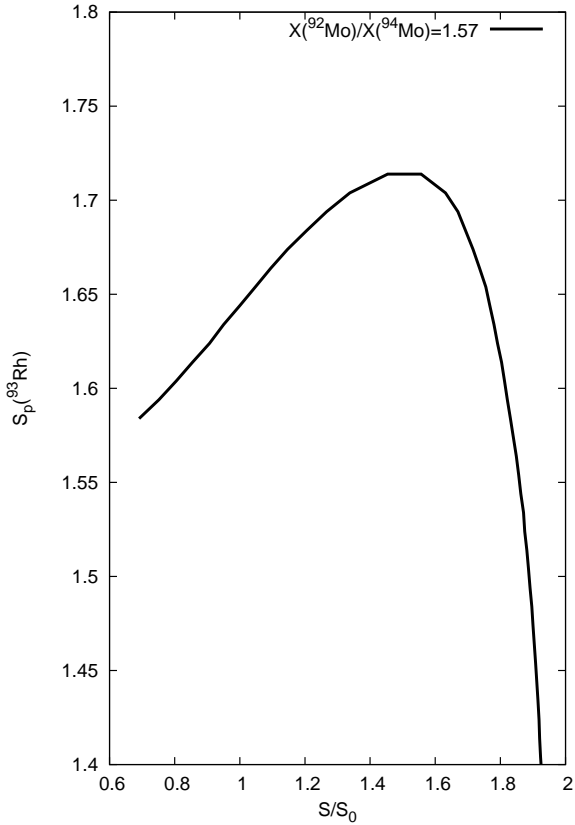}
\caption{The allowed values of $S_P(^{93}\textrm{Rh})$ as a function 
of changes in entropy, $S$ relative to the entropy of ``trajectory 6'', 
$S_0 = 77$,  in the outflowing wind for the solar ratio of 
$^{92}\textrm{Mo}/^{94}\textrm{Mo}$.} \label{enfig}
\end{figure}

We systematically investigated the effect relevant nuclear uncertainties on 
this reaction flow using the model described in \cite{Pruet06,Pruet05}. 
We find that variation within current uncertainties \cite{Audi03b} of the $^{91}$Rh proton separation 
energy and the $^{92}$Rh proton separation energy does not change the 
solar abundance ratio of $^{92}$Mo to$^{94}$Mo whereas the ratio is 
highly sensitive to the proton separation energy of $^{93}$Rh. 
Fig.~\ref{enfig} shows the dependence of the solar ratio $^{92}$Mo to $^{94}$Mo 
to variations in entropy of ``trajectory 6''. We show that 
$S_p(^{93}\textrm{Rh})=1.63$ MeV is a solution to a range of entropy 
variations between 0.8 and 1.6 of the nominal value. The figure also shows no 
solution above $S_p(^{93}\textrm{Rh})=1.71$ MeV. 

\begin{figure}
\includegraphics[width=0.48\textwidth]{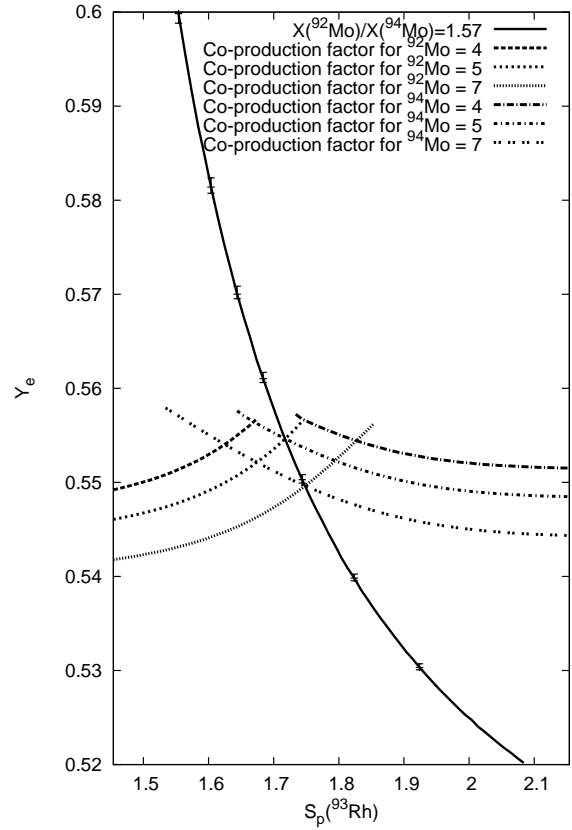}
\caption{The solid line shows the solution for $Y_e$ and 
$S_P(^{93}\textrm{Rh})$ where the 
$^{92}\textrm{Mo}/^{94}\textrm{Mo}$ ratio in the outgoing wind 
matches the solar ratio. Error bars indicate the extent of similar 
lines for ratios of 1.54 and 1.59. Also shown are the solutions where 
$^{92}\textrm{Mo}$ and $^{94}\textrm{Mo}$ are coproduced within a factor 
4, 5, and 7. A solution is found for a co-production factor of 5 with 
$Y_e$=0.555 and $S_P(^{93}\textrm{Rh})$ = 1.72 (see main text for details).
\label{yefig}}
\end{figure}

Fig.~\ref{yefig} shows the dependence of the solar ratio 
$^{92}$Mo to$^{94}$Mo to variations in entropy in ``trajectory 6'' 
as a function of $Y_e$ and $S_p(^{93}\textrm{Rh})$. 
The figure also shows the solutions where $^{92}$Mo and $^{94}$Mo are 
co-produced within a factor 4,5 and 7. Isotopes produced with precisely the 
solar abundance pattern have equal production factors. A co-production 
factor of no more than 7 is typically regarded as acceptable as the global 
characteristics of nucleosynthesis are sensitive to details of the outflow. 

The conclusion that the $^{92}$Mo and $^{94}$Mo ratio is predominantly influenced 
by $S_p(^{93}\textrm{Rh})$ has been shown to be robust (Fisker et al., 
submitted for publication). However, our calculations predict that  
$S_p(^{93}\textrm{Rh})=1.63$ MeV whereas recent experimental results suggest that  
$S_p(^{93}\textrm{Rh})=2.0001\pm 0.008$ MeV \cite{efw08}. This leads to the tentative conclusion 
that proton rich winds under the conditions in the model investigated here can 
not be the sole source of the solar $^{92}$Mo and $^{94}$Mo.

\section{The contribution from neutron rich pockets}
Using the same supernova model as above, the contribution to core-collapse 
nucleosyntheis of the proton-rich bubbles and proton-rich winds was investigated in \cite{Pruet05,Pruet06} 
However, some bubbles also contains neutron-rich matter that is ejected in coincidence with the proton-rich bubbles. Here,
investigate their contribution to the overall nucleosynthesis by considering 
newly extrated trajectories with $0.47 \leq Y_e \leq 0.50$.

For $Y_e$ closer to 0.5, primarily $^{56,57,58}$Ni are formed. The flow from 
these nuclei leads to $^{64}$Ge. Unlike the $\nu p$-process \cite{Froehlich06a}, 
there is not a sufficient amount of protons left at this time for neutrinos provide 
sufficient numbers of neutrons to capture on $^{64}$Ge and thus move beyond this waiting 
point. As a result, heavier isotopes are not co-produced with the 
${}^{62}$Ni and ${}^{64}$Zn isotopes. In particular, there is 
no overproduction of the light p-nuclei for $Y_e\leq 0.5$. For $Y_e$ closer to 0.47, 
primarily ${}^{58,59,60}$Ni are formed. This means that the ${}^{64}$Ge waiting point 
is circumvented which leads to overproduction of ${}^{74}$Se, ${}^{78}$Kr, 
and ${}^{92}$Mo which is co-produced with ${}^{64}$Zn. With decreasing $Y_e$, 
${}^{92}$Mo production falls off and the overproduction of N=50 nuclei ensues.

\begin{figure}
\includegraphics[width=0.48\textwidth]{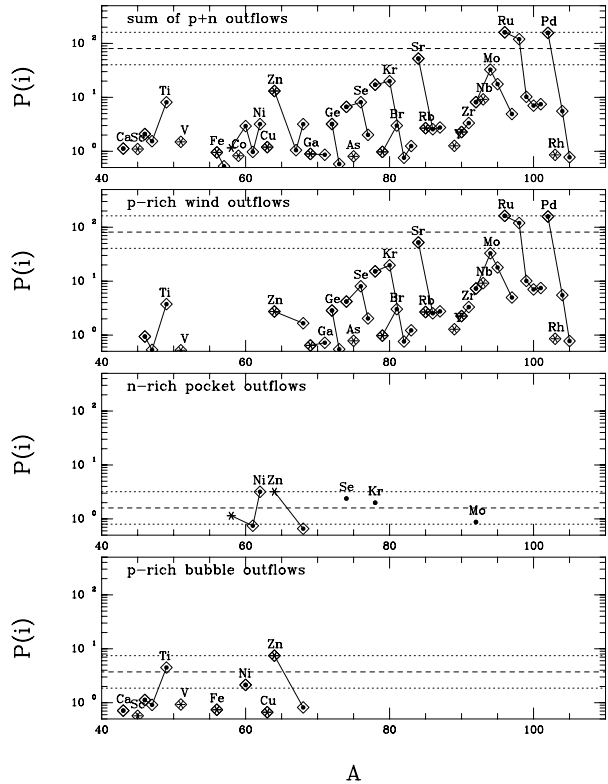}
\caption{Production factors of the neutron-rich trajectories of the convective 
bubble ejecta. The most abundant isotope for a given element is shown with an asterisk. 
Diamonds indicate that the isotope was made primarily as a radioactive 
progenitor.}
\end{figure}

The figure shows the integrated production factors for all studied neutron-rich bubble trajectories. 
The most produced isotopes in the neutron-rich parts of the bubble relative 
to solar abundances are ${}^{62}$Ni and ${}^{64}$Zn which originate in bubbles 
with Ye closer to 0.5. These are co-produced along with ${}^{74}$Se and 
${}^{78}$Kr which originate in the bubbles with Ye closer to 0.47. The 
neutron-rich bubbles add ${}^{74}$Se, ${}^{78}$Kr, and ${}^{92}$Mo to the 
bubble-outflow, but this contribution is much smaller than the contribution from the 
proton-rich winds when neutrino interactions are included. The neutron-rich 
bubbles also add ${}^{62}$Ni and ${}^{64}$Zn to the total outflow but only in 
comparable amounts to the wind outflows and the proton-rich bubble outflows. 
Our results show that the overproduction factors of the neutron-rich bubbles 
folded with the mass-ejecta does not contribute significantly to the nucleosynthesis 
of the light p-nuceli compared to the nucleosynthesis of the proton-rich material. 

\begin{theacknowledgments}

This work was performed under the auspices of the U.S. Department
of Energy by Lawrence Livermore National Laboratory in part under
Contract W-7405-Eng-48 and in part under Contract DE-AC52-07NA27344.
It was also supported, in part, by the SciDAC Program of
the US Department of Energy (DC-FC02-01ER41176).
The project in Garching
was supported by the Deutsche Forschungsgemeinschaft
through the Transregional Collaborative Research Centers SFB/TR~27
``Neutrinos and Beyond'' and SFB/TR~7 ``Gravitational Wave Astronomy'',
and the Cluster of Excellence EXC~153
``Origin and Structure of the Universe''.
The SN simulations were performed on the national supercomputer
NEC SX-8 at the High Performance Computing Center Stuttgart (HLRS)
under grant number SuperN/12758.

\end{theacknowledgments}

\bibliographystyle{aipproc}   

\end{document}